\documentstyle[11pt,newpasp,twoside,epsf]{article}
\begin{document}
\title{Tracing the relation between black holes and dark haloes}
\author{Pieter Buyle, Maarten Baes \& Herwig Dejonghe}
\affil{Astronomical Observatory Ghent University, Krijgslaan 281 S9, 9000 Ghent, Belgium, Pieter.Buyle@UGent.be}
\begin{abstract}
We present new velocity dispersion measurements for a set of 12 spiral
galaxies and use them to derive a more accurate $v_c-\sigma$ relation
which holds for a wide morphological range of galaxies. Combined with
the $M_{BH}-\sigma$ relation, this relation can be used as a tool to
estimate supermassive black hole (SMBH) masses by means of the
asymptotic circular velocity. Together with the Tully-Fisher relation,
it serves as a constraint for galaxy formation and evolution models.
\end{abstract}
During the past few years, fiducial mass estimates for the putative
SMBHs in the centre of galaxies have become available. These SMBH
masses were found to correlate closely to various properties of their
host galaxies (Kormendy et al. 1995; Ferrarese et al. 2000; Gebhardt
et al. 2000; Graham et al. 2001). Recently, Ferrarese (2002)
discovered a tight correlation between the central velocity dispersion
and the asymptotic circular velocity measured well into the dark
matter dominated region of spiral galaxies. Strikingly, this
correlation seems to be satisfied not only by spiral galaxies but also
by ellipticals. This indicates a link between the formation and
evolution of SMBHs and dark matter haloes.\vspace{0.4cm}\newline We
took long-slit spectra of 12 galaxies with the EFOSC2 spectrograph on
the ESO 3.6~m telescope. We derived the central velocity dispersions
of the galaxies from the Mg\rm{I} and Fe lines around 520~nm. The
asymptotic circular velocities were taken from Palunas \& Williams
(2000).\vspace{0.4cm}
\newline Using linear regression on our data combined with those of
Ferrarese (2002) we find the correlation
\begin{equation}
\log{\left(\frac{v_c}{u_0}\right)}=(0.96\pm 0.11)\log{\left(\frac{\sigma}{u_0}\right)}+(0.21\pm 0.02)\ \ \ {\mathrm with}\ u_0=200\ {\mathrm km\,s^{-1},}
\end{equation}
with a negligible intrinsic scatter of $\chi_r^2=0.281$ (see
figure). We confirm that this relation holds over a large
morphological range from late-type spirals to massive ellipticals, and
that it appears to break down in the low mass regime ($\sigma<80$
km\,s$^{-1}$). Combining the $v_c-\sigma$ relation with the equally
tight correlation that links the velocity dispersions and SMBH masses
we obtain
\begin{equation}
\log{\left(\frac{M_{BH}}{M_{\sun}}\right)}=(4.21\pm 0.60)\log{\left(\frac{v_c}{u_0}\right)}+(7.24\pm 0.17)\ \mathrm{,}
\end{equation}
where the most up-to-date $M_{BH}-\sigma$ relation (Tremaine et
al. 2002) was used. This expression can serve as an easy tool to
estimate the masses of SMBHs. As the asymptotic circular velocity is a
measure for the total gravitational mass, the previous relation can be
transformed into a relation between the masses of the SMBH and the
dark halo. Considering state-of-the-art high-resolution CDM
cosmological simulations (Bullock et al. 2001), we establish a
non-linear link
\begin{equation}
\frac{M_{BH}}{10^8\ M_{\sun}}\sim 0.11\left(\frac{M_{DM}}{10^{12}\ M_{\sun}}\right)^{1.27}\ \mathrm{.}
\end{equation}
\newline
In conclusion, the $v_c-\sigma$ relation together with the
$M_{BH}-\sigma$ and Tully-Fisher relations clearly points towards an
intimate interplay between the various components of the galaxies and
constitutes a strong test for galaxy formation and evolution models.
\begin{figure}
\plottwo{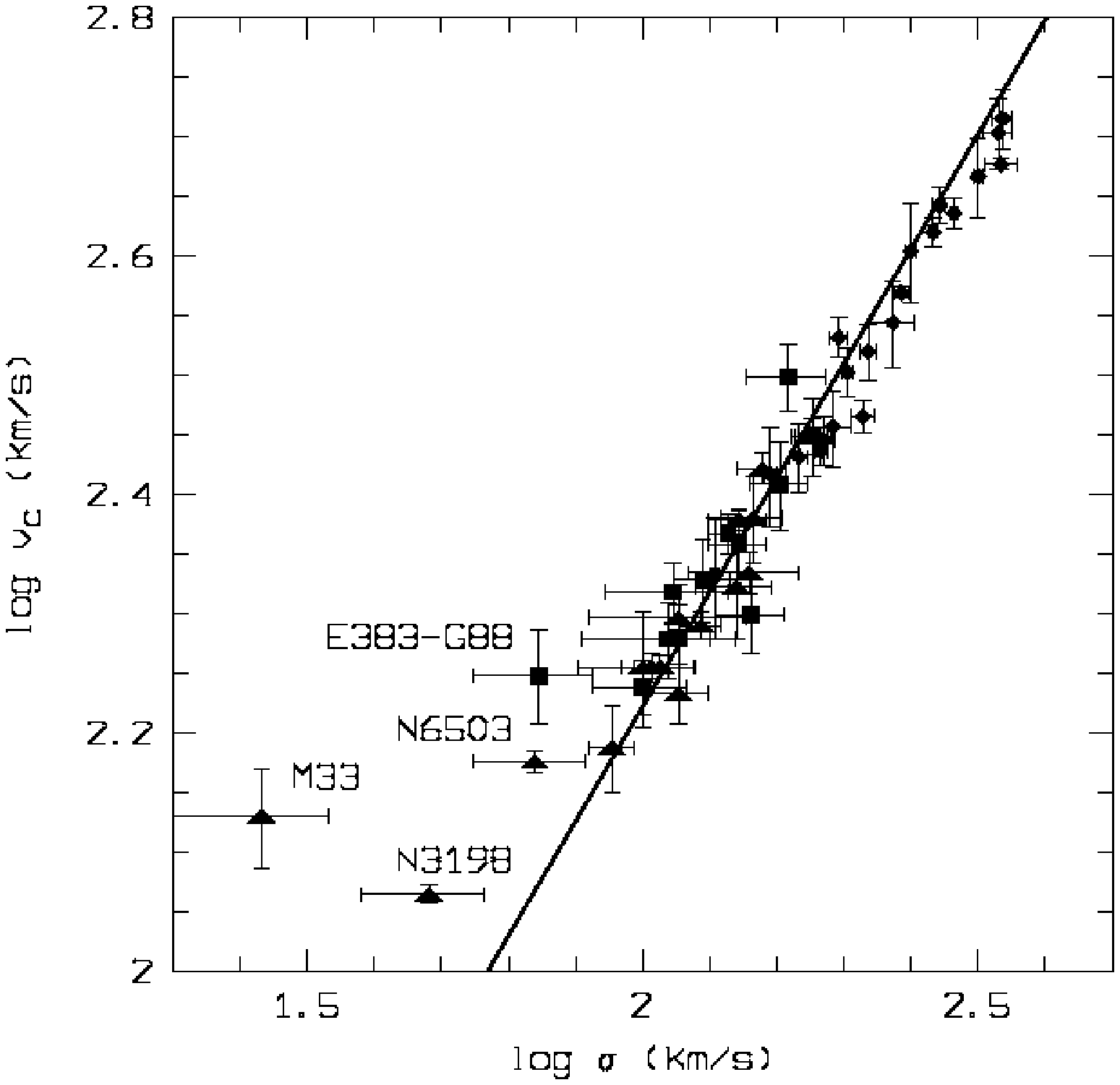}{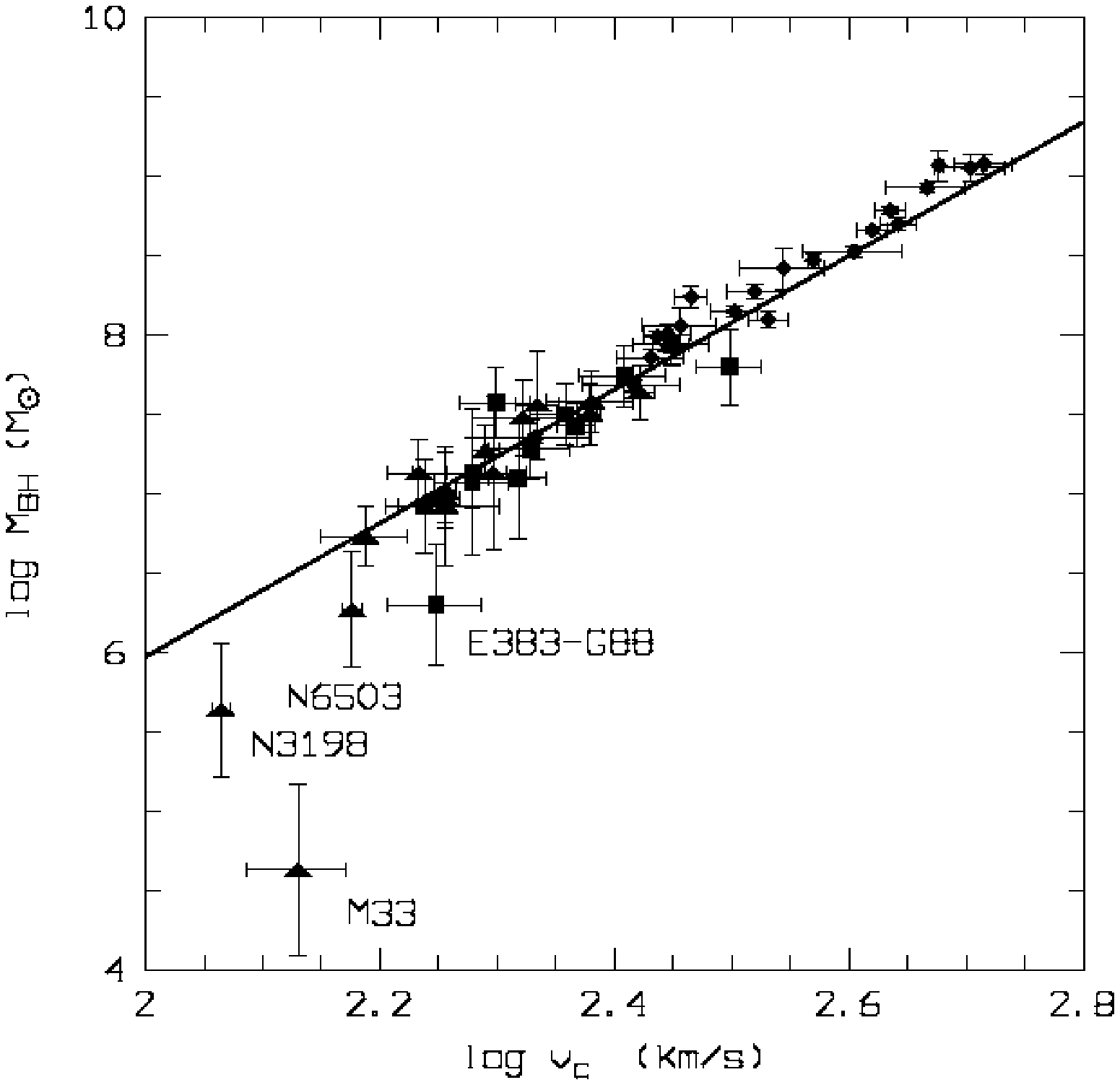}
\caption{The left and right figure show respectively equations (1) and (2). Ferrarese's data (2002) are represented by triangles, Kro-\\ nawitter's (2000) by circles and ours by squares. Both relations appear to break down for $\sigma < 80$ km\,s$^{-1}$ or $v_c<150$ km\,s$^{-1}$.}
\end{figure})

\end{document}